\documentclass[%
 reprint,
 amsmath,amssymb,
 aps,
 pra,
]{revtex4-2}

\usepackage{graphicx}
\usepackage{dcolumn}
\usepackage{bm}
\usepackage{xcolor}
\usepackage[normalem]{ulem}

\begin{document}

\preprint{APS/123-QED}

\title{Intensity correlations in the forward four-wave mixing driven by a single pump}

\author{A. A. C. de Almeida}
 \email{alexandre.cavalcantialmeid@ufpe.br}
\author{M. R. L. da Motta}
\author{S. S. Vianna}%
\affiliation{Departamento de F\'{i}sica, Universidade Federal de Pernambuco, 50670-901, Recife, Pernambuco, Brazil}

\date{\today}

\begin{abstract}
We study the field intensity fluctuations of two independent four-wave mixing signals generated in a cold rubidium sample as well as the transmission signals. We employ an experimental setup using a single CW laser to induce the nonlinear process in a forward geometry using either parallel and circular or orthogonal and linear polarizations of the input fields. Even though the spectra of each experimental configuration are significantly different due to the distinct level structures of each scenario, both cases present intensity–intensity cross-correlation of the four-wave mixing signals. We also calculate the cross-correlation between the input fields and draft a theoretical model that points that resonant phase-noise to amplitude-noise conversion allows the observation of Rabi oscillations in the cross-correlation curves.
\end{abstract}

\maketitle


\section{\label{sec:level1}Introduction}

For the past three decades, there has been an interest in studying light fluctuations when it interacts with matter. The pioneer experimental work of Yabusaki \textit{et al} \cite{yabusaki1991} showed that one may obtain spectroscopic information using the intensity fluctuations of a laser beam interacting with rubidium vapor. In the sequence, R. Walser and P. Zoller \cite{walser1994} provided a theoretical framework to explain this new type of spectroscopy and especially how the conversion of phase-noise to amplitude-noise is at the root of this phenomenon. Ever since these works, a large amount of research has been produced studying these fluctuations in the light-matter interaction, with interesting results such as the study of correlations and anti-correlations in electromagnetically induced transparency \cite{martinelli2004, cruz2007, xiao2009, florez2013}, controlling intensity noise correlations and squeezing of four-wave mixing processes via polarization \cite{li2016}, and the generation of correlated and anticorrelated fields via atomic spin coherence \cite{yang2012}.  

There have been different approaches to the problems regarding these fluctuations,  concerning whether the analysis is in the frequency domain or time domain. In this last approach, there is a set of works led by the group of M. O. Scully \cite{sautenkov2005, varzhapetyan2009, ariunbold2010}, which are of interest to the problem we present here. Our experiment uses a similar setup with a single cw laser, however, along with the correlations between transmitted beams, we also investigate the correlations between two non-linear signals generated by two independent FWM processes.

In this sense, we present experimental observations of strong correlations between intensity fluctuations of two FWM signals generated through the interaction of laser light with a cold rubidium sample. The correlation between the input laser fields is also detailed, with observations that agree with the literature. We compare different polarization configurations, which access distinct internal energy level structures. Since we use a cold atomic sample, the system has a narrow Maxwell-Boltzmann distribution, so we can study how the correlations behave as a function of the laser detuning. This is an advantage of the cold system if compared to an atomic vapor, in which several velocity groups can respond to the input laser even if one changes the detuning, as long as it is inside the Doppler broadening range.

Furthermore, we observe an oscillatory behavior compatible with Rabi oscillations \cite{papoyan2021} in the correlation functions. The intriguing feature is that we can detect these oscillations long after the transient period, retrieving the frequency information through the correlation function. This idea of extracting an oscillation frequency using the correlation function has been used in other contexts such as the observation of quantum beats in spontaneous emission \cite{norris2010} or the observation of temporal beats in Raman Stokes fields \cite{chen2010}. The theoretical model we build supports the idea that it is only because of the fluctuations that we can detect this oscillatory behavior of the system. 

The paper is organized as follows: In Sec. II, we detail the experiment and all the experimental results. In particular, we show the time series of all four signals and the corresponding second-order correlation functions. Moreover, we demonstrate how the correlation changes regarding variations in the intensity and frequency of the input fields. Sec. III is devoted to building a simple theoretical model that can provide insight into the physical meaning of the results. We conclude by summarizing the relevant achievements of this work in Sec. IV.

\section{\label{sec:level2}Experimental setup and results}

In the experiment, we use a single cw laser to generate two input laser beams labeled by their wave-vectors $\vec{k}_a$ and $\vec{k}_b$, as presented in Fig. 1. These two beams interact with a cold atomic sample of $^{87}$Rb atoms in a magneto-optical trap (MOT), typically with $10^8-10^9$ atoms cooled to temperatures of hundreds of $\mu$K. In the experimental configuration with linear and orthogonal polarization, the input fields are aligned with the atomic cloud using a polarizing beam-splitter (PBS). In the other case, with circular and parallel polarization, we substitute the two PBS for beam-splitters and add two quarter-wave plates before and after the atomic cloud. 

\begin{figure}
\centering
\includegraphics[width=1\linewidth]{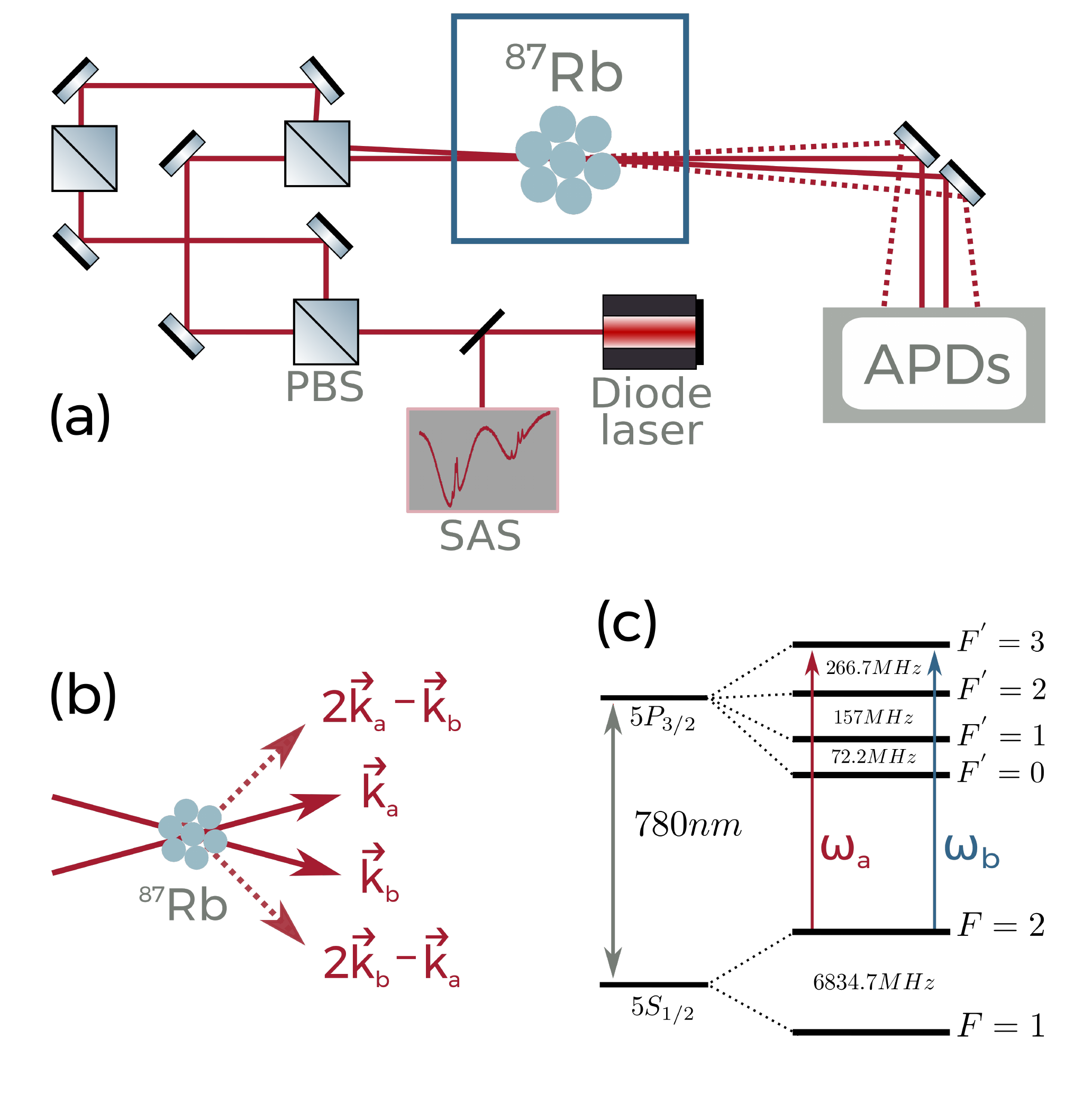}
\caption{(a) Simplified scheme of the experimental setup; (b) Wave-vectors of the four signals (two FWM and two transmissions); (c) Hyperfine structure of the $D_2$ line of $^{87}$Rb.}
\label{fig1}
\end{figure}

We are interested in the two FWM signals generated in directions $2\vec{k}_{a}-\vec{k}_{b}$ and $2\vec{k}_{b}-\vec{k}_{a}$, as shown in Fig. \ref{fig1}(b). Therefore, we investigate processes in which two photons of one of the beams are absorbed, and one photon of the other beam is emitted, generating new coherent signals. The input beams are in an almost copropagating configuration, with a small angle of 10 mrad between them to allow the spatial separation of all four signals. This type of forward geometry is challenging since scattered light from one beam might arrive at the detection position of the other beams. We detect the two FWM signals and the transmissions of the input beams $E_a$ and $E_b$ with avalanche photodiodes (APD) of the model APD120A/M from \textit{Thorlabs}.

The beams that induce the degenerate FWM processes are tuned near the closed transition $\left| F=2 \right\rangle$ $\rightarrow$ $\left| F'=3 \right\rangle$, of the $D_2$ line of $^{87}$Rb (see Fig. \ref{fig1}(c)). Since this is the same transition excited by the cooling laser of the MOT, we use a temporal scheme to generate and acquire the signal of interest. We shut down for 2 ms all trapping fields, that is, the cooling laser and the anti-Helmholtz coils. This time interval is enough so that the atoms cannot gain much speed and therefore move away from the center of the MOT. After turning off the trapping fields, there is a 20 $\mu$s delay, allowing the repump laser to properly prepare the atoms in the ground state $\left| F=2 \right\rangle$, where they can interact with the FWM inducing laser. The repump laser is always active throughout the measurement to guarantee the proper state preparation.

We lock the frequency of the input laser to a saturation absorption spectroscopy peak and tune the laser frequency using an Acoustic-Optical Modulator (AOM). In the time interval in which the MOT fields are off, we acquire data from a time series with typically 100 $\mu$s intervals of all four signals. In these measurements, one must be careful with the detuning with respect to the resonance because for small detunings, or high laser intensity, the radiation pressure can disperse the atomic cloud. Therefore, there is a practical limitation in our experiment, since the FWM signal increases when the input laser is closer to resonance and for higher laser intensities, the same regime that increases the radiation pressure.

Given this limitation, the experimental configuration with linear and orthogonal polarization has an advantage. The spectra, shown in Fig. 2(a), is wider than the spectra for the circular and parallel case (see Fig 2(b)). Furthermore, it has a dip around the resonance so that the maximum signal is slightly off-resonance. The fundamental difference between each case is that the linear and orthogonal polarization interacts with a sum of $\Lambda$ systems in the form of the Zeeman sublevels, while the circular case is modeled by a pure two-level system. The presence of two degenerate ground levels together with equally powerful input laser beams that scan their frequency simultaneously induces a coherent population trapping that prevents the signal from being generated on resonance \cite{orriols79, agrawal81, boublil91}.

\begin{figure}
\centering
\includegraphics[width=1\linewidth]{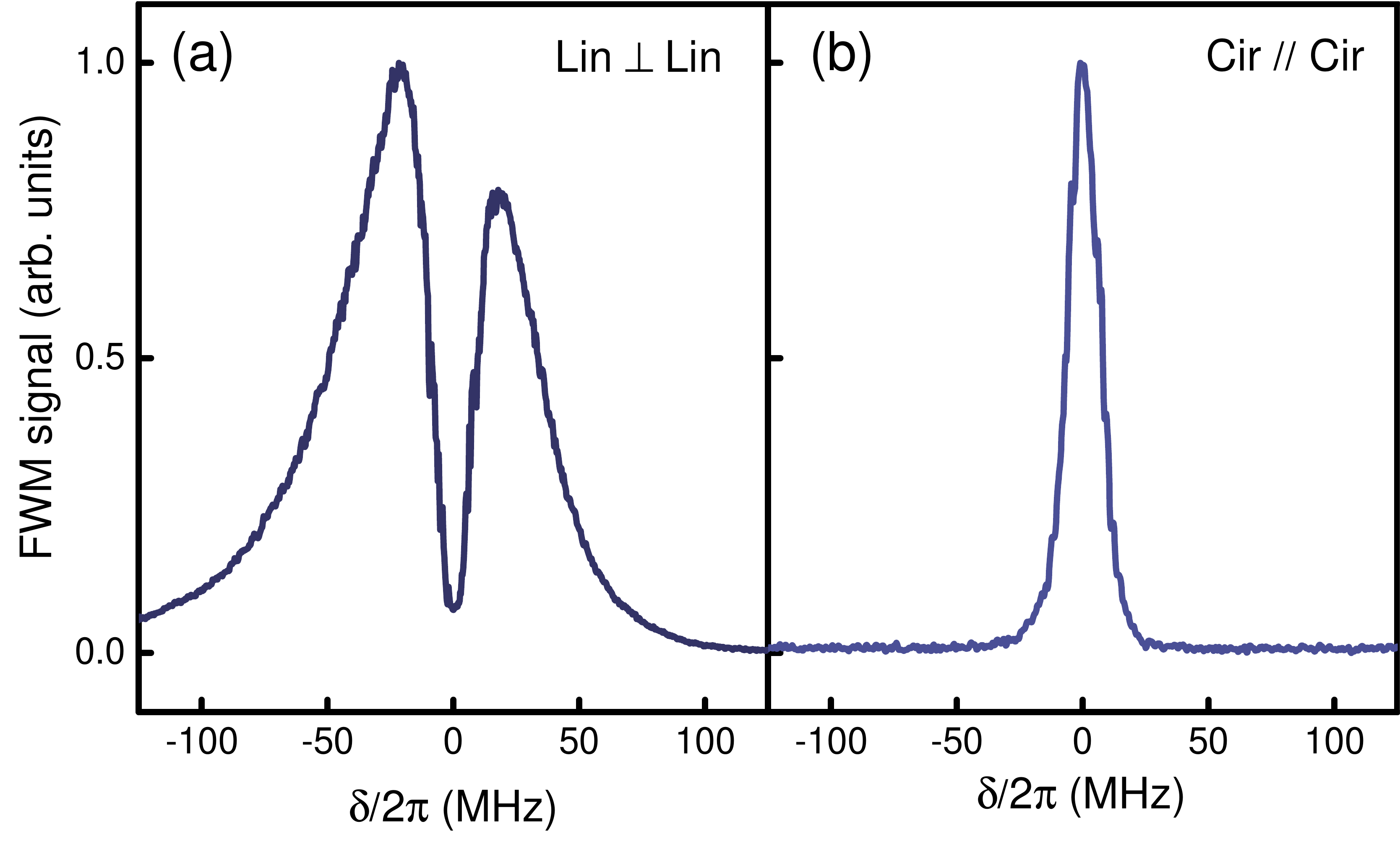}
\caption{FWM spectra with input laser intensity $I_a=I_b=10$ mW/cm$^{2}$ for (a) linear and orthogonal polarizations; (b) circular and parallel polarizations.}
\label{fig2}
\end{figure}

The time series of the intensity fluctuations of all four signals and for the two polarizations are presented in Fig. 3. The presented data is part of the 100 $\mu$s recorded time series and has been filtered with high-pass ideal FFT filter with a cutoff frequency of 500 kHz, to eliminate any slow fluctuations of the signals. In each figure, 3(a) for circular and parallel polarization and 3(b) for orthogonal and linear polarization, we show the intensity fluctuations versus time of the two FWM signals (red and blue lines) for an input laser intensity of $I_a=I_b=3.3$ mW/cm$^{2}$ and a detuning from the excited state of $\delta/2\pi = 70$ MHz. It is noticeable that fluctuations behave similarly, even though they are not identical. Due to the experimental difficulties regarding the radiation pressure, one can only achieve a signal with a good signal-to-noise ratio far from resonance and with input lasers with intensities close to or above the saturation intensity.

\begin{figure}
\centering
\includegraphics[width=1\linewidth]{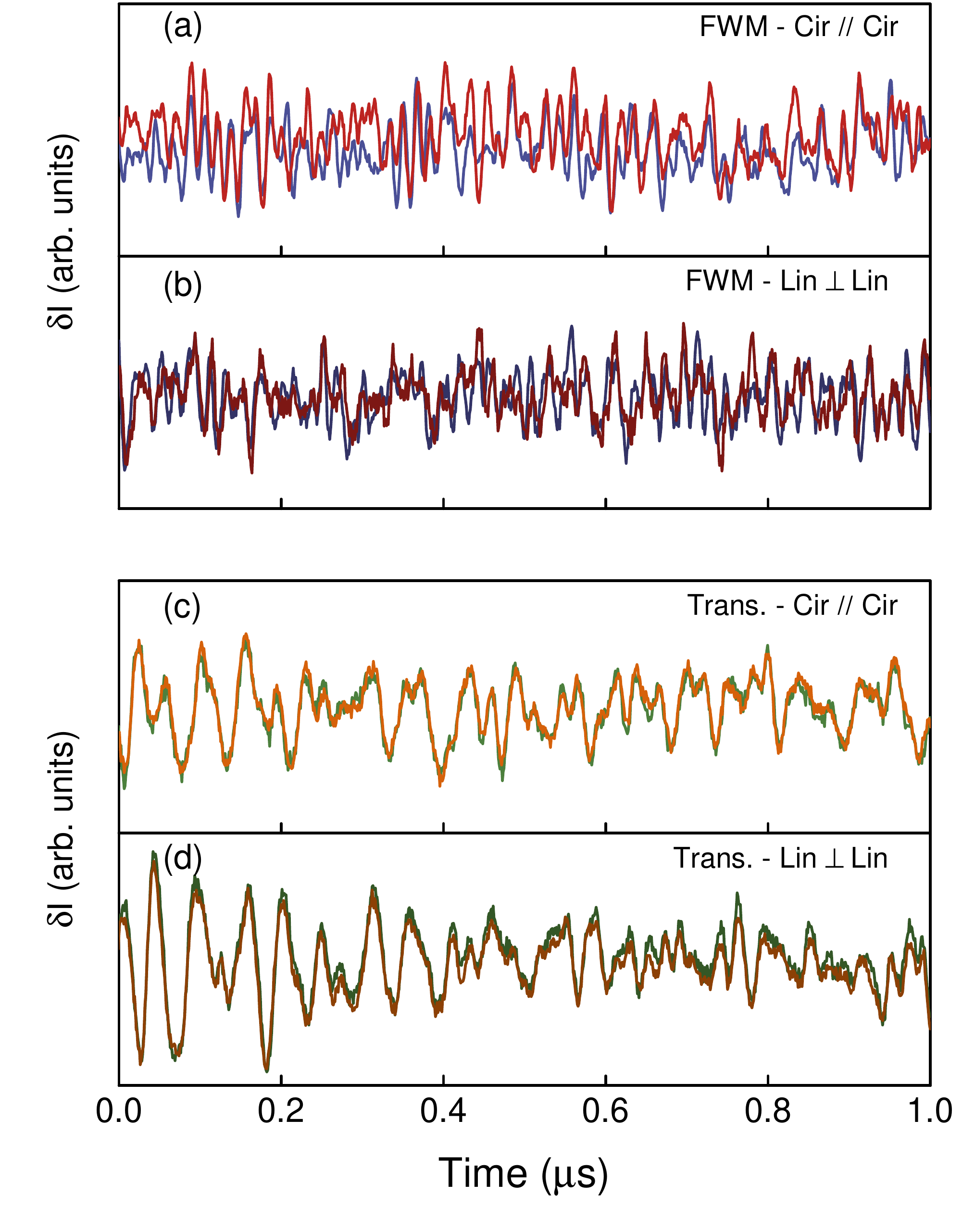}
\caption{Time series of the intensity fluctuations for the FWM signal with input laser intensity of $I_a=I_b=3.3$ mW/cm$^{2}$, detuning from the excited state of $\delta/2\pi = 70$ MHz and (a) circular and parallel polarizations; (b) linear and orthogonal polarizations. Time series of the intensity fluctuations for the transmittance of the input lasers with $I_a=I_b=0.15$ mW/cm$^{2}$, $\delta/2\pi = 15$ MHz and (c) circular and parallel polarizations; (d) linear and orthogonal polarizations. }
\label{fig3}
\end{figure}

On the other hand, the intensity fluctuations of the input lasers can be obtained a lot closer to resonance as long as the intensity is small. In Figs. 3(c) and (d) we show the time series of the input lasers (orange and green lines) for a input laser intensity of $I_a=I_b=0.15$ mW/cm$^{2}$ and a detuning from the excited state of $\delta/2\pi = 15$ MHz. It is clear that these results are remarkably synchronized and should present near-perfect correlations, a known result \cite{ariunbold2010}.

These correlations can be quantified with the second-order correlation function $G_{ij}^{(2)}(\tau)$ \cite{sautenkov2005, varzhapetyan2009, ariunbold2010, yang2012} for intensity fluctuations of two optical beams with time delay $\tau $. It is given by

\begin{equation}
G_{ij}^{(2)}(\tau)=\frac{\left\langle \delta I_{i}\left(t\right)\delta I_{j}\left(t+\tau\right)\right\rangle }{\sqrt{\left\langle \delta I_{i}\left(t\right)^{2}\right\rangle \left\langle \delta I_{j}\left(t+\tau\right)^{2}\right\rangle }},
\label{g2}
\end{equation}

\noindent
where the $\delta I_{i,j} (t) = I_{i,j} (t) - \left\langle I_{i,j} (t)\right\rangle$ are the time-dependent intensity fluctuations with $ \left\langle I_{i,j} (t)\right\rangle$  being the average intensities of the laser fields and $i,j = a, b, s1, s2$  the labels to designate the two input fields and the two FWM signals, respectively. 

\begin{figure}
\centering
\includegraphics[width=1\linewidth]{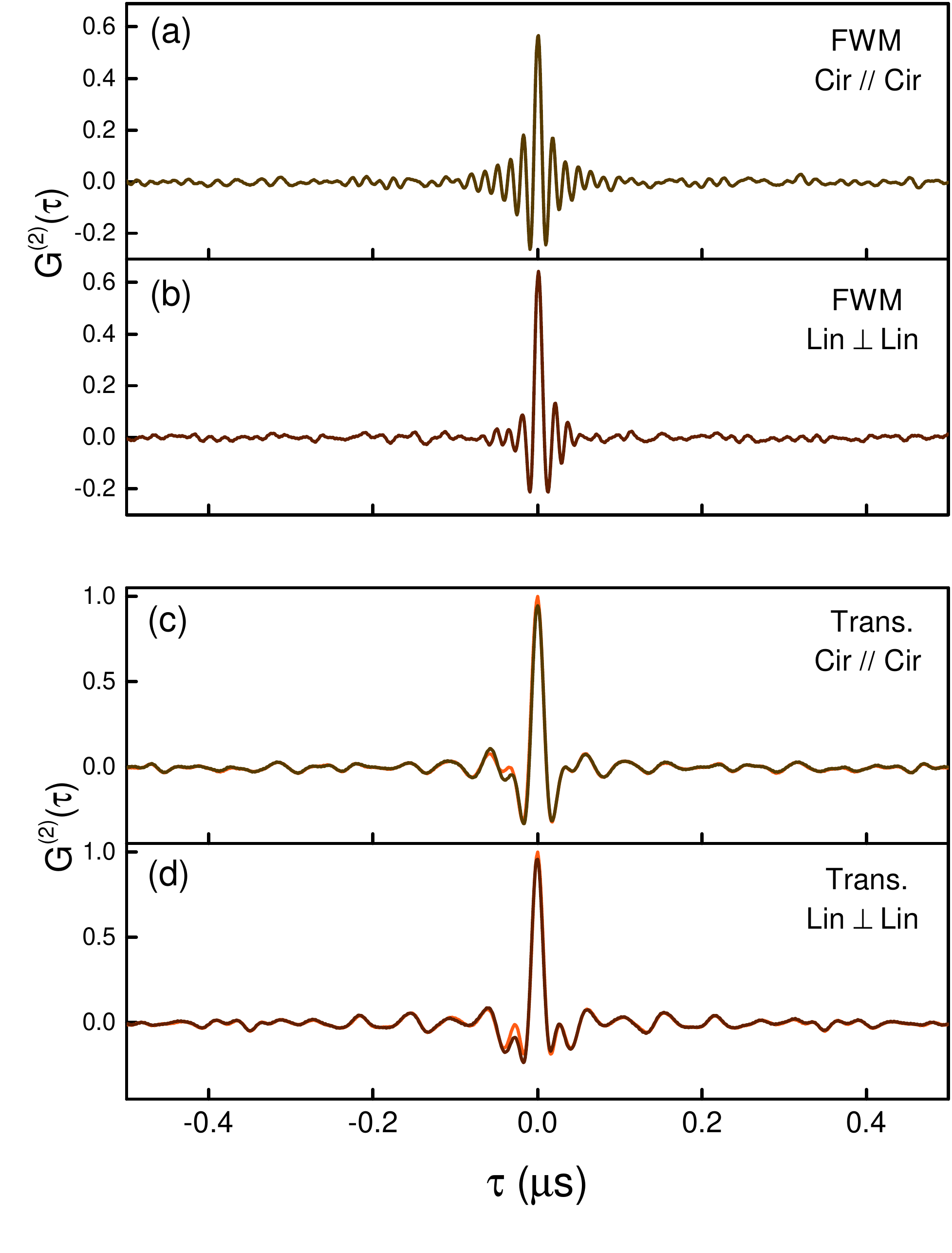}
\caption{Second-order correlation function $G_{ij}^{(2)}(\tau)$ between the FWM signals with input laser intensity of $I_a=I_b=3.3$ mW/cm$^{2}$, detuning from the excited state of $\delta/2\pi = 70$ MHz and  (a)  circular and parallel polarizations; (b) linear and orthogonal polarizations. Second-order correlation function $G_{ij}^{(2)}(\tau)$ for the transmittance of the input lasers (dark brown line) and autocorrelation (orange line) with $I_a=I_b=0.15$ mW/cm$^{2}$, $\delta/2\pi = 15$ MHz and (c) circular and parallel polarizations; (d) linear and orthogonal polarizations. }
\label{fig4}
\end{figure}

We present the intensity fluctuations correlation functions $G_{ij}^{(2)}(\tau)$ for the pairs of time series of Fig. 3 in Fig. 4. These correlation functions have peaks at zero time delay with amplitudes (Pearson coefficient) of $\approx 0.6$ for the FWM signals and over 0.95 for the transmission signals. This confirms the expectation of Fig. 3 that there is a strong temporal positive correlation in the intensity fluctuations of the output signals. Moreover, in Figs. 4(c) and (d) we also present the autocorrelation (orange curve) for the intensity fluctuations of the $a$ laser beam. These curves are remarkably similar to the cross-correlation (dark brown curves) of the two transmission signals, specially concerning the oscillations near zero delay.

The cross-correlation we observe in the transmission beams, arises due to the resonant phase-noise to amplitude-noise conversion \cite{mcintyre1993, walser1994, camparo1997, camparo1998, camparo1999}. The resonant interaction with atoms plays a critical role in this result. If there were no atoms, or if the input laser was not near resonance, there would be no correlation. The point we raise here is that this conversion also happens to the FWM signals, creating correlated fields, even though they come from processes that cannot occur simultaneously for the same atom. Ultimately, the fields are correlated because they all come from the same laser with the same phase fluctuations. In the next section, we build a simple model that provides insight on how the phase fluctuations of the input laser manifest themselves in the detected signals.

Given this argument, it would be interesting to look at the correlations between one of the input laser fields and one of the FWM signals. However, the experiment limits this situation as there is not a suitable choice of parameters to obtain both signals simultaneously. To achieve the maximum FWM signal, one must increase the intensity of the input laser as much as the radiation pressure on the atomic sample allows. On the other hand, to measure the intensity fluctuations of the input laser, it cannot have a high intensity, otherwise, the medium will saturate, and most of the detected photons will not interact with the atomic cloud.

\begin{figure}
\centering
\includegraphics[width=1\linewidth]{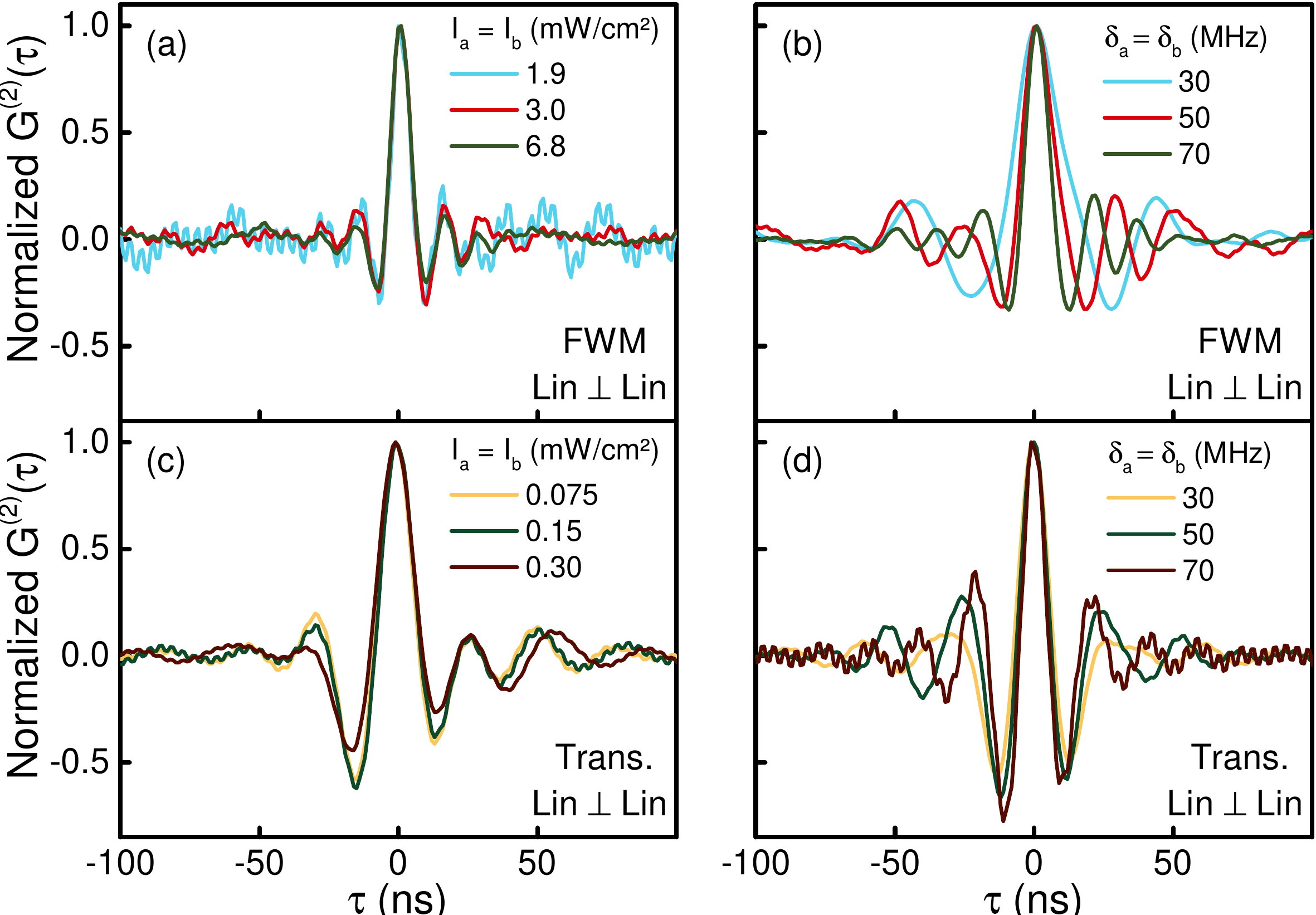}
\caption{Normalized second-order correlation function $G_{ij}^{(2)}(\tau)$  between intensity fluctuations of FWM signals with linear and orthogonal polarization (a) varying input laser intensity with $\delta/2\pi = 85$ MHz  and (b) varying input laser detuning; (c) Transmission signals with linear and orthogonal polarization varying input laser intensity with $\delta/2\pi = 25$ MHz  and (d) varying input laser detuning}
\label{fig5}
\end{figure}

An intriguing feature of these correlation curves is that the width of the FWM correlation peak seems different from the transmission correlation peak, which is to be expected since they have different intensities and detunings. In Ref. \cite{sautenkov2005}, a fairly similar experiment but using an atomic vapor and a magnetic field to break the degeneracy of the Zeeman sublevels, the authors comment that the widths of the correlation peaks are associated with a power broadening of the single photon resonance in the Rb vapor. Therefore, we repeated the time series for each pair in different intensities for a fixed detuning. These results are presented in Fig. \ref{fig5}(a) for the FWM signals and \ref{fig5}(c) for the transmission beams, both with linear orthogonal polarization, with normalized correlation functions to achieve a proper comparison of the widths.

In most cases, we increased the input laser intensity by a maximum factor of 3, and the width of the correlation peak did not change significantly. We believe it is because the atomic medium is not truly saturated in any of the measurements, either when we compare transmission or FWM signals. In the case of the FWM signals, the input laser fields are above saturation intensity, but the nonlinear signal itself is weak. 

These results indicate that the second-order correlation function behaves differently regarding changes in the detuning. To be able to investigate this is an advantage of using a cold sample instead of an atomic vapor. In hot systems, the Doppler broadening is significant, meaning that variations of the detuning inside the Maxwell-Boltzmann curve will always find a resonant velocity group.

In Figs. \ref{fig5}(b) and \ref{fig5}(d), we present the correlation between transmission signals and between FWM signals, in the experimental configuration with linear polarization, for three different detunings. The results are similar for circular polarization. The most noticeable feature of these results is that correlation curves do get wider as the frequency of the input laser approaches the resonance. Moreover, far from resonance, an oscillation of the correlation curves becomes clearer and with higher frequency. There are regions of correlation, $G_{ij}^{(2)}(\tau)>0$, and, regions of anti-correlation, $G_{ij}^{(2)}(\tau)<0$. This behavior was already apparent in the previous correlation curves for the FWM signals (see Fig. \ref{fig4}(a) and \ref{fig4}(b)), as they were all far from resonance. It is important to remark that the results of Fig. \ref{fig5}(d) are for a fixed input intensity of $0.15$ mW/cm$^2$, while the results for the FWM signals, in Fig. \ref{fig5}(b), are for different intensities in each case. However, if one looks at the generalized Rabi frequency $\tilde{\Omega}=\sqrt{\Omega^{2}+\delta^{2}}$, it is approximately equal to the detuning since the Rabi frequency of the input beams is close to the natural linewidth of the transition. We must vary the intensity in these measurements to maximize the FWM signal in each measurement, otherwise, the signal-to-noise ratio would not allow proper visualization of the correlation.

Since to obtain the correlation between intensity fluctuations of the transmission signals, we can work with very low intensities, it is easy to tune the laser frequency without pushing away the atoms, making it possible to achieve a detailed map of the correlation as a function of detuning. We present such a map in Fig. 6. In this graph the broadening of the central peak near resonance becomes clearer. Furthermore, it is noticeable the presence of oscillations near the central peak. It seems that the frequency of this oscillation gets smaller near resonance. A Fourier analysis of the curves in Fig. 6 shows (see Fig. 7) that they have a spectral component compatible with $\tilde{\Omega}$. 

\begin{figure}
\centering
\includegraphics[width=1\linewidth]{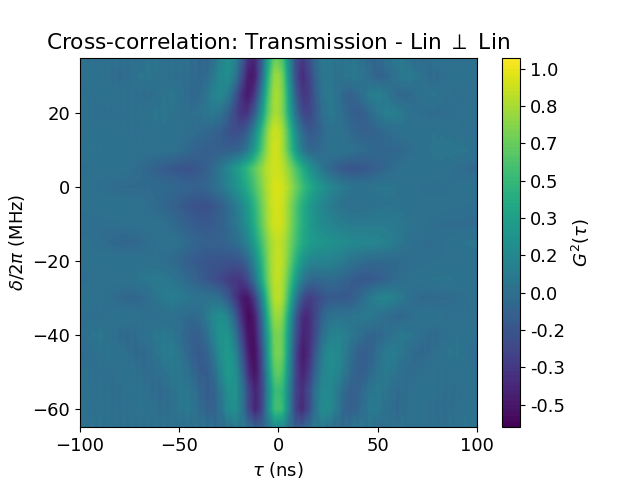}
\caption{Second-order correlation function $G_{ij}^{(2)}(\tau)$ as a function of the detuning $\delta/2 \pi$ between transmission signals with linear and orthogonal polarization. The input laser intensity is $I_{a}=I_{b}=$0.15 mW/cm$^2$.}
\label{fig6}
\end{figure}

Therefore, these oscillations in the second-order correlation function are connected to the generalized Rabi frequency of the input laser. This indicates that in the conversion process between phase fluctuations of the laser into intensity fluctuations through the interaction with the atomic medium, the intensity fluctuations oscillate with approximately the generalized Rabi frequency \cite{papoyan2021}. 

One could expect to see this oscillation in the raw data, that is, in the time series of Fig. 3. However, they are not noticeable in this case as these measurements are taken long after the transient period when this oscillation should be more noticeable. Furthermore, the signal we acquire is the average signal of the light emitted by the atomic ensemble and not by a single atom. In fact, a Fourier analysis of that data does not reveal any spectral component in particular. 

On the other hand, a higher-order measurement should be able to retrieve the spectral information of the system \cite{chen2010, norris2010}. This is possible with the intensity fluctuations correlation function $G_{ij}^{(2)}(\tau)$, which does present a noticeable spectral component, as we already mentioned. The Fourier analysis of these curves (Fig. 4, 5, and 6) shows that there is a spectral component compatible with the generalized Rabi frequency, or since the laser intensity is usually small, compatible with the detuning. It is expected that this is an approximate result because even for a simple two-level system, the presence of spontaneous emission decay modifies how the temporal solution of the optical Bloch equations oscillates, but the values should be close to $\tilde{\Omega}$.

To verify this claim, we plot in Fig. 7 the spectral component present in each correlation curve (red dots) compared with the absolute value of the detuning (solid line). There is a reasonable agreement between the two results, supporting our argument. One can see in the transmission case (see Fig. 7(c) and (d)) that near resonance, the spectral component of the correlation curves moves away from the detuning as the Rabi frequency gets more relevant to the generalized Rabi frequency.

\begin{figure}
\centering
\includegraphics[width=1\linewidth]{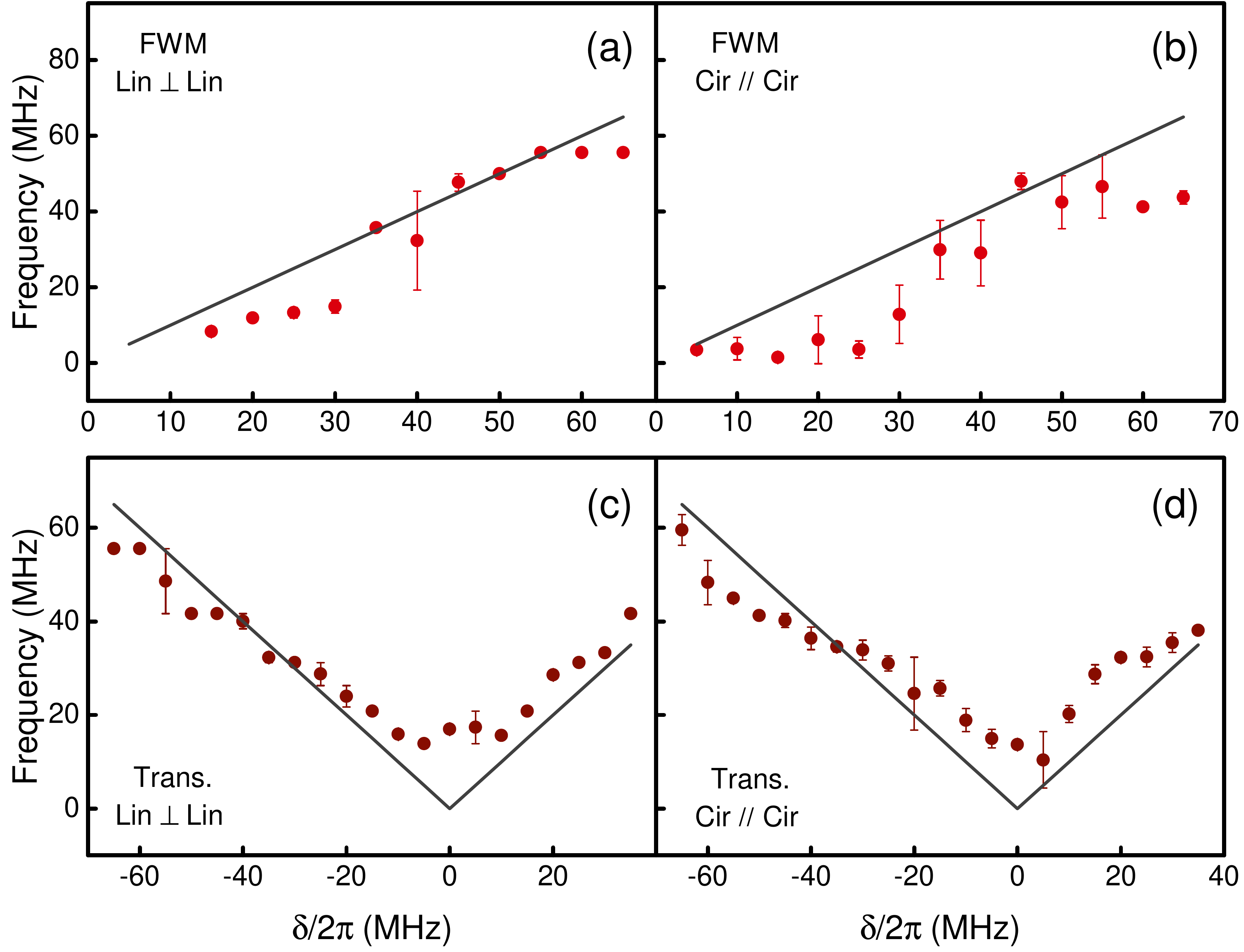}
\caption{Frequency of the oscillation in the correlation function for FWM signals with (a) linear and orthogonal polarizations and (b) circular and parallel polarizations; Transmission signals with (c) linear and orthogonal polarizations and (d) circular and parallel polarizations. The solid lines are the absolute value of the detuning.}
\label{fig7}
\end{figure}

One final comment on the result of the correlation between FWM signals is related to why we observe a positive correlation and not a competition between signals, that is, an anti-correlation. Yang \textit{et al} \cite{yang2012} observed an anti-correlation between FWM signals in an atomic vapor using a $\Lambda$-sytem. This situation compares to our results using linear and perpendicular configurations. Our case differs from theirs because the ground states are degenerate, rendering a symmetrical system that forbids competition between the fields. The results of Ref. \cite{ariunbold2010} show how this degeneracy controls the correlation by introducing an external magnetic field that can break the degeneracy, and change from perfect correlation to anti-correlation.

\section{\label{sec:level3} Theoretical model}

We employ the model of Ref. \cite{ariunbold2010} to explain the main observed features in the correlation results between the transmission signals. We extend the results of this previous work by exploring the dependence of the correlation on the detuning, with special attention to the presence of the Rabi oscillations. Our system allows this analysis since the Doppler broadening can be neglected in cold atomic systems. Therefore, we begin modeling the experimental results with linear and perpendicular polarization configuration, which is connected to a three-level system. To model the features of the circular and parallel polarization scenario, we will use the same set of equations but eliminate one of the ground states. 

The treatment of the problem begins considering an electric dipole coupling as the interaction Hamiltonian 

\begin{equation}
\hat{H}_{int} = -\hbar\sum^3_{j \neq k}\left(\Omega_{l}+c.c.\right)\left|j\right\rangle\left\langle k\right|,
\end{equation}

\noindent
where $\Omega_{l}=\frac{\mu_{jk}E_{l}}{2\hbar}$ ($l = a$ or $b$) is the Rabi frequency with $\mu_{jk}$ being the transition dipole moment and $E_{l}$ the electric fields. In the linear and perpendicular case these fields are represented by

\begin{equation}
    \begin{split}
        \vec{E}_{a}&=\frac{1}{2}\left[\varepsilon_{a}(t)e^{-i\left(\omega_{a}t+\phi\left(t\right)-k_{a}z\right)}+c.c.\right]\frac{\left(\hat{\sigma}^{+}+\hat{\sigma}^{-}\right)}{\sqrt{2}};\\\vec{E}_{b}&=\frac{1}{2}\left[\varepsilon_{b}(t)e^{-i\left(\omega_{b}t+\phi\left(t\right)-k_{b}z\right)}+c.c.\right]\frac{\left(i\hat{\sigma}^{+}-i\hat{\sigma}^{-}\right)}{\sqrt{2}}.
    \end{split}
\end{equation}

\noindent
where $\varepsilon_{l}$ is the amplitude of the electric field; $\omega_{l}$ is the optical frequency, $\phi(t)$ is the fluctuating phase, and $\vec{k}_{l}$ is the associated wave-vector. The polarization vector is represented in the circular base as it highlights how these fields interact with the $\Lambda$-system.

We consider that the electric fields have a fluctuating phase $\phi(t)$, described by a Wiener-Levy diffusion process \cite{agarwal76}. For these processes, the average of the stochastic variable is zero and the average of the two-time correlation is given by $\overline{\left\langle \dot{\phi}\left(t\right)\dot{\phi}\left(t'\right)\right\rangle }=2D\delta\left(t-t'\right)$; where $D$ is the diffusion coefficient. Often in the literature, the stochastic process chosen to represent this phase is the Ornstein–Uhlenbeck \cite{gardiner2004quantum, orzag}, which includes an extra term to the Wiener process with an exponential function of the time delay.
    
Moreover, we introduce an extra simplification to the model: we consider that the field $a$ is in one of the transitions while the field $b$ is in the other, as Fig. 8 shows. That is, we only take one circular component of each field. We do so to allow each one-photon coherence to oscillate with the frequency of one of the input fields. Hence, the number of coupled Stochastical Differential Equations (SDE) decreases significantly. A complete treatment of this system, including even wave-mixing processes of superior orders, is performed in Ref. \cite{wilsongordon2018}. In this model, they perform a Floquet expansion of the density-matrix elements in the frequency of the input fields and their combinations. One could include a stochastical phase in this last model, but it would take the system from nine coupled SDEs as we write it in this work to a few tens of equations. The numerical solution becomes unstable and demanding in terms of computation time.

\begin{figure}
\centering
\includegraphics[width=.5\linewidth]{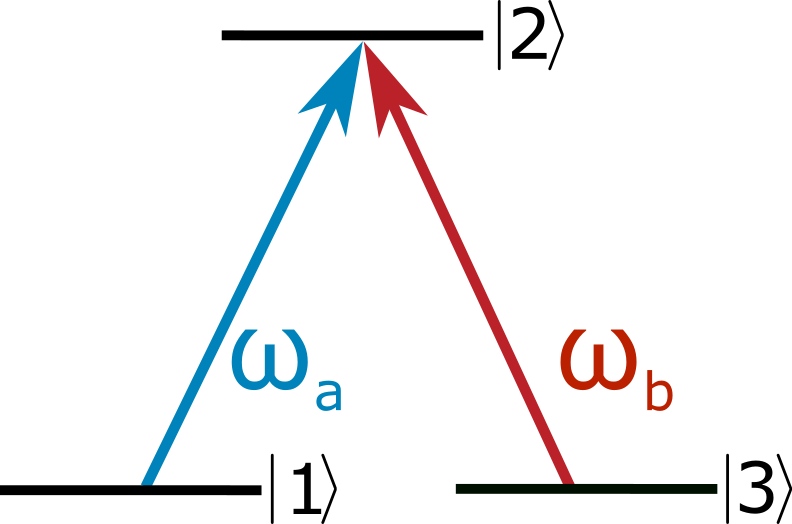}
\caption{Simplified level scheme for the configuration with linear and perpendicular polarization.}
\label{fig8}
\end{figure}

Using this Hamiltonian and considering the approximations we described, it is possible to write Liouville's equation

\begin{equation}
\frac{\partial\rho_{jk}}{\partial t} = -(i\omega_{jk} + \gamma_{jk})\rho_{jk} - \frac{i}{\hbar}\left\langle j \right| [ \hat{H}_{int},\hat{\rho} ] \left| k \right\rangle,\\
\end{equation}

\noindent
where $\gamma_{jk}$ is the decay rate of the density matrix element $\rho_{jk}$, and $\omega_{jk}$ is the frequency of the $\left|j\right\rangle$ $\rightarrow$ $\left|k\right\rangle$ transition. The Bloch equations in the rotating-wave approximation can be written as:

\begin{equation}
\begin{aligned}
\dot{\rho}_{11}&=-i\sigma_{12}\Omega_{a}+i\sigma_{21}\Omega_{a}^{*}+\Gamma_{21}\rho_{22};\\\dot{\rho}_{22}&=i\sigma_{12}\Omega_{a}-i\sigma_{21}\Omega_{a}^{*}-i\sigma_{23}\Omega_{b}^{*}+i\sigma_{32}\Omega_{b}-(\Gamma_{21}+\Gamma_{23})\rho_{22};\\\dot{\rho}_{33}&=i\sigma_{23}\Omega_{b}^{*}-i\sigma_{32}\Omega_{b}+\Gamma_{23}\rho_{22};\\\dot{\sigma}_{12}&=-\sigma_{12}\left(i\delta_{a}+\gamma_{12}-\dot{\phi}\left(t\right)\right)-i(\rho_{11}-\rho_{22})\Omega_{a}^{*}-i\sigma_{13}\Omega_{b}^{*};\\\dot{\sigma}_{13}&=-\sigma_{13}\left(i\delta_{a}-i\delta_{b}+\gamma_{13}\right)-i\sigma_{12}\Omega_{b}+i\sigma_{23}\Omega_{a}^{*};\\\dot{\sigma}_{32}&=-\sigma_{32}\left(i\delta_{b}+\gamma_{23}-\dot{\phi}\left(t\right)\right)-i(\rho_{33}-\rho_{22})\Omega_{b}^{*}-i\sigma_{31}\Omega_{a}^{*}.
\end{aligned}
\end{equation}

\noindent
The $\sigma_{jk}$ terms are the coherence from equation (4) in the rotating frame, whereas $\Gamma_{jk}$ are the decay rates of the populations. The missing coherence equations are the complex conjugate of the ones presented.

Since the set of Eqs. (5) contains stochastical terms, we must solve them numerically using Itô's calculus. As previously mentioned, we take a typical stationary stochastic process, the Ornstein-Uhlenbeck process, to describe the phase fluctuations. This process satisfies the SDE:

\begin{equation}
    dX_{t}=\alpha\left(\gamma-X_{t}\right)dt+\beta dW_{t}
\end{equation}

\noindent
where the Itô's diffusive process $dX_{t}$ has a deterministic part and a stochastic one. The deterministic term, the first one, has a magnitude of the mean drift $\alpha$ while the asymptotic mean is $\gamma$. If $X_t>\gamma$ the drift will be negative and the process will go towards the mean. If $X_t<\gamma$ then the opposite happens, the drift is positive and the process moves away from the mean. As for the stochastic part, it is a Brownian motion $W_t$ with a magnitude constant $\beta$.

We solve the system of SDEs using a stochastic Runge-Kutta for scalar noise algorithm. This algorithm possesses a good accuracy for our problem, with a thin distribution of residuals. We also use the same Brownian increment $dW_t$ for both one-photon coherences, as the original fluctuation comes from a single laser. Finally, we probed several choices of parameters of the Ornstein-Uhlenbeck process, but the outcomes are not drastically different as long as the variance of the process, given by $\beta^2 / 2\alpha$, is small.

Once the numerical simulation is complete, we have access to a theoretical time series of all elements of the density-matrix. Therefore, for each of these terms, we can calculate the second-order correlation function for the intensity fluctuations. However, we must establish the link between the density-matrix elements and the actual detected signal. To do so, we solve the wave equation derived from Maxwell's equations, neglecting the transverse derivatives of the electric field. With a few algebraic manipulations and using the adiabatic approximation we can obtain the simple differential equation $\frac{\partial\Omega_{l}}{\partial z}=i\kappa_{2j}\sigma_{2j}$, where $\kappa_{2j}=\frac{\omega_{l}N\mu_{2j}^{2}}{2\hbar\epsilon_{0}c}$, with $N$ being the number of atoms.

Solving this equation leads to the fields we detect in the experiment after they propagate in the sample. To do so, we use the fact that MOT diameter $L$ is much smaller than the Rayleigh length of the fields in play. Therefore, it is adequate to consider the thin-medium regime, which implies that we can make use of the equations in Ref. \cite{ariunbold2010} and rewrite the second-order correlation function of the intensity fluctuations of Eq. (1) as:

\begin{equation}
G^{\left(2\right)}\left(\tau\right)=\frac{\left\langle \mathrm{Im}\left(\delta\sigma_{21}\left(t\right)\right)\mathrm{Im}\left(\delta\sigma_{23}\left(t+\tau\right)\right)\right\rangle }{\sqrt{\left\langle \left[\mathrm{Im}\left(\delta\sigma_{21}\left(t\right)\right)\right]^{2}\right\rangle \left\langle \left[\mathrm{Im}\left(\delta\sigma_{23}\left(t\right)\right)\right]^{2}\right\rangle }}.
\end{equation}

\noindent
To obtain this last result, we neglected second-order terms when calculating the field intensity.

Notably, in the model, as it is, only the transmission results can be directly explored. However, as our numerical solution renders the elements of the density-matrix in all orders, the nonlinear effects of wave-mixing are also built in the one-photon coherence. Naturally, the stronger term should be the lower-order one, which is indeed connected to the transmission.

\section{\label{sec:level4}Theoretical results}

After solving the system of coupled SDEs, we have a numerical simulation of the time series for the transmission signals. An example of a single realization of such series is presented in Fig. 9 (a) for a detuning of $\delta/2 \pi = 30$ MHz and a Rabi frequency $\Omega_a=\Omega_b=0.1 \Gamma$. We simulate the signal for 20 $\mu$s but neglect the first half of the series to ensure that the transient period is not present. Employing Eq. (7), we calculate the second-order correlation function for the series of Fig. 9(a) and present the results in Fig. 9(b).

\begin{figure}
\centering
\includegraphics[width=1\linewidth]{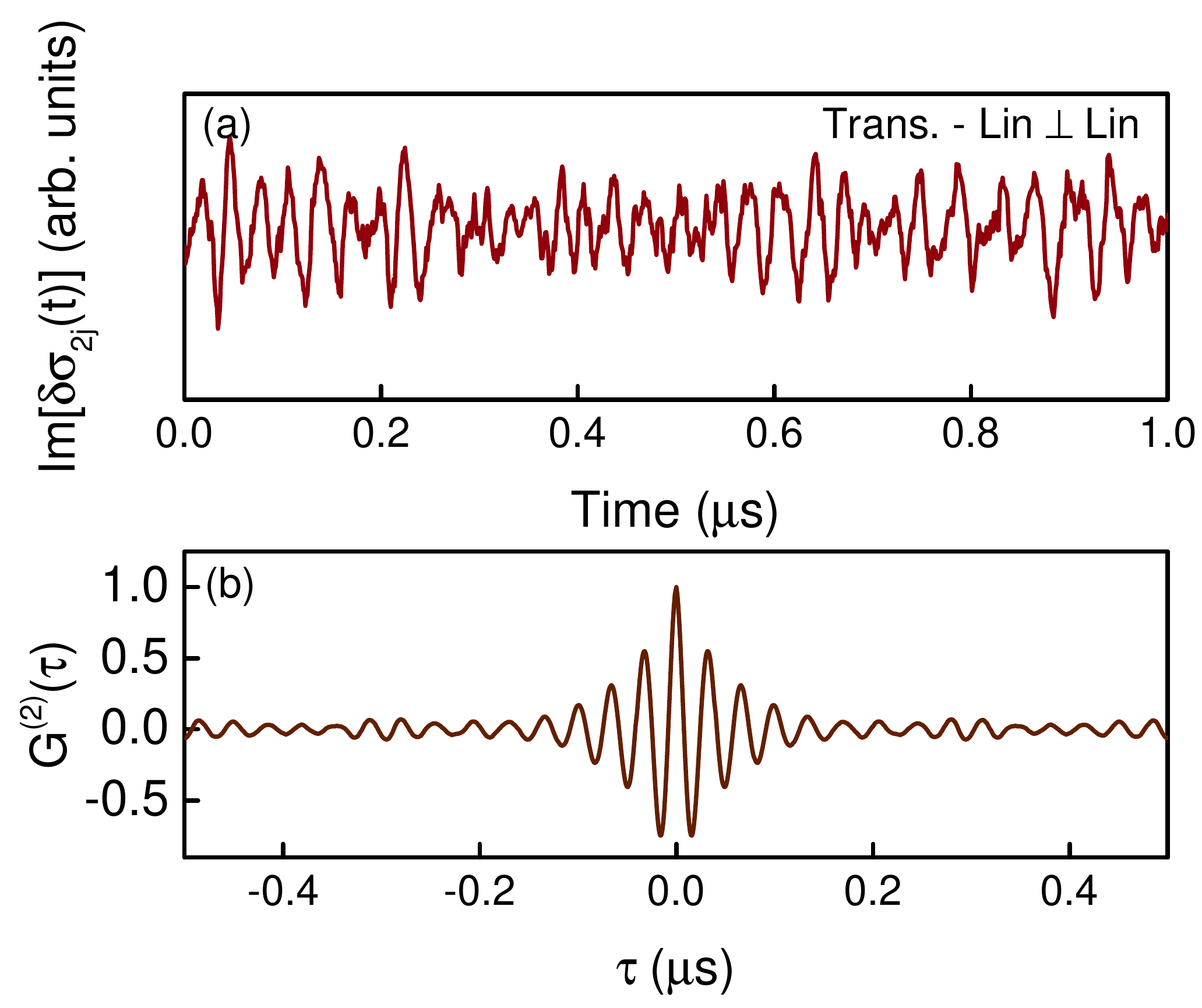}
\caption{(a) Numerical simulation of a time series of the intensity fluctuations for the transmission signals with input Rabi frequency $\Omega_a=\Omega_b=0.1 \Gamma$, detuning from the excited state of $\delta/2\pi = 30$ MHz and linear and orthogonal polarizations. (b) Second-order correlation function $G_{ij}^{(2)}(\tau)$ between the time series of (a).}
\label{fig9}
\end{figure}

Since we use the same Brownian increment $dW_t$ for both signals, they must be perfectly correlated as in Fig. 9(b). It is easy to introduce different increments for each input field and control how large their correlation is by stating that $dW_t^{(1)}=dW_t^{(2)}+\sqrt{1-\rho^2}dW_t^{(3)}$, where $\rho$ ranges from zero to one; that is, one increment is equal to the other with the addition of a third increment. However, we use a single cw laser, so the stochastic phase each input field carries should be the same.

We presented the case of cross-correlation between input field intensity fluctuations for the scenario with linear and perpendicular polarization. If we eliminate one of the ground states and, therefore, all the equations and terms connected to it in Eqs. (5), then we would have a two-level system. This is the scenario with circular and parallel polarization. However, as the results are remarkably similar to those in Fig. 9, we do not present them here.

\begin{figure}
\centering
\includegraphics[width=1\linewidth]{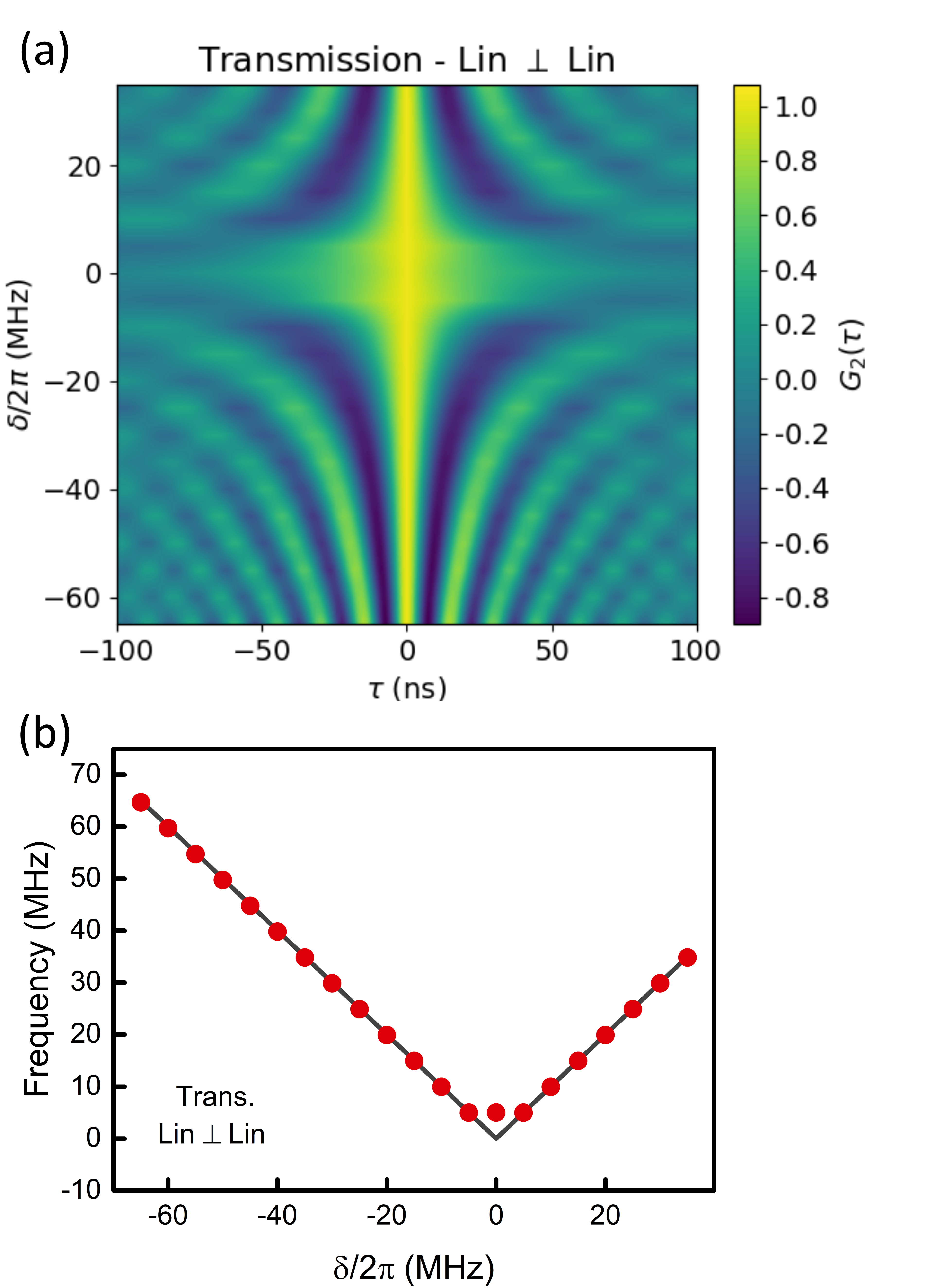}
\caption{(a) Second-order correlation function $G_{ij}^{(2)}(\tau)$ as a function of the detuning $\delta/2 \pi$ between theoretical transmission signals with linear and orthogonal polarization. The input laser Rabi frequency is $\Omega_{a}=\Omega_{b}=0.1 \Gamma$. (b) Frequency of the oscillation in the correlation function of (a). The solid line is the absolute value of the detuning}
\end{figure}

A theoretical map such as the one presented in Fig. 6 can be achieved and it is presented in Fig. 10(a). It shows a broadening of the correlation peak compatible with the behavior of the experimental result. Moreover, the theoretical results highlight the oscillation of the correlation curve in the generalized Rabi frequency. 

A graph similar to the one shown in Fig. 7 is presented in Fig. 10(b) for the transmission signals in the linear and perpendicular polarization case. This graph agrees with the experimental results and therefore supports that the frequency of the oscillation we see in the correlation curves is indeed well described by the generalized Rabi frequency.

We must emphasize that our results were able to reveal this oscillating behavior because the experiments were performed in a cold atom cloud. In a vapor cell, for example, this signature would have been washed away due to the atomic movement. The Doppler integration should change the observation of these oscillations, as the correlation curve would contain the response of several velocity groups.

\section{\label{sec:level5}Conclusions}

We have successfully demonstrated that there are temporal correlations between intensity fluctuations of two distinct degenerate FWM signals in a cold rubidium sample. It is noteworthy that this is the first demonstration of these correlations in degenerate FWM processes, which due to this degeneracy, do not present competitive signals and therefore have a positive correlation between them. In a complementary scenario to ours are the results of Ref. \cite{yang2012} that presents an anticorrelation between FWM signals due to the nondegeneracy of the ground states. 

Furthermore, since our cold atomic system allows a proper definition of detuning, namely, there is not a significant Doppler broadening, we can study how the correlation between FWM signals and between transmission signals behave as a function of detuning. The results show that the system exhibit Rabi oscillations that can be revealed by the second-order correlation function long after the transient. The theoretical model from Ref. \cite{ariunbold2010} was used to provide numerical results for the transmission signals that support the experimental findings. Even though the model only deals with the transmission signals, it provides a good insight that the mechanism behind the correlations, and the Rabi oscillations we see in them, is the conversion of phase-noise to amplitude-noise due to the interaction of the laser with the atoms. The FWM signals should follow a very similar behavior, so we believe that the results regarding the cross-correlation between transmission signals can be extrapolated to the nonlinear case.

\begin{acknowledgments}
This work was supported by CAPES (PROEX 534/2018, No.
23038.003382/2018-39). The authors would like to acknowledge the financial support from Brazilian agencies CAPES and CNPq.  A. A. C. de Almeida acknowledges financial support by CNPq (141103/2019-1). M. R. L. da Motta acknowledges financial support from CNPq (130306/2020-7) and CAPES (88887.623521/2021-00).
\end{acknowledgments}

\bibliography{biblio}

\end{document}